%% file: main.tex
\documentclass[twoside]{article}

\usepackage{aistats2019, times}
\input{math_commands.tex}

\usepackage{graphicx}
\usepackage{overpic}
\usepackage{algorithm}
\usepackage{algpseudocode}
\usepackage{float} 

\usepackage{array}
\usepackage{xcolor}
\usepackage{hyperref}
\usepackage{url}
\usepackage{amsthm}
\usepackage{amssymb}
\usepackage{comment}

\usepackage[round]{natbib}
\renewcommand{\bibname}{References}
\renewcommand{\bibsection}{\subsubsection*{\bibname}}

\begin{document}

%

%

\twocolumn[

\aistatstitle{Active Learning over DNN: Automated Engineering Design Optimization for Fluid Dynamics Based on Self-Simulated Dataset}

\aistatsauthor{ Yang Chen* }

\aistatsaddress{Cranbrook Schools} 
]

\begin{abstract}
   \input{abstract}
\end{abstract}
\input{introduction}
\input{methodology}
\input{results}
\input{discussion}
\input{conclusion}

\input{Acknowledgments}

\bibliography{iclr2019_conference}
\bibliographystyle{iclr2019_conference}

\end{document}

%% file: math_commands.tex
\usepackage{amsmath,amsfonts,bm}

\def\eqref#1{equation~\ref{#1}}

\def\1{\bm{1}}

\DeclareMathAlphabet{\mathsfit}{\encodingdefault}{\sfdefault}{m}{sl}
\SetMathAlphabet{\mathsfit}{bold}{\encodingdefault}{\sfdefault}{bx}{n}



%% file: abstract.tex
   Optimizing fluid-dynamic performance is an important engineering task. Traditionally, experts design shapes based on empirical estimations and verify them through expensive experiments. This costly process, both in terms of time and space, may only explore a limited number of shapes and lead to sub-optimal designs. In this research, a test-proven deep learning architecture is applied to predict the performance under various restrictions and search for better shapes by optimizing the learned prediction function. The major challenge is the vast amount of data points Deep Neural Network (DNN) demands, which is improvident to simulate. To remedy this drawback, a Frequentist active learning is used to explore regions of the output space that DNN predicts promising. This operation  reduces the number of data samples demanded from ~8000 \citep{eismann2017optimization} to 625. The final stage, a user interface, made the model capable of optimizing with given user input of minimum area and viscosity. Flood fill is used to define a boundary area function so that the optimal shape does not bypass the minimum area. Stochastic Gradient Langevin Dynamics (SGLD) is employed to make sure the ultimate shape is optimized while circumventing the required area. Jointly, shapes with extremely low drags are found explored by a practical user interface with no human domain knowledge and modest computation overhead. 

%% file: introduction.tex
\section{Introduction}


Fluid Mechanics is a resource consuming process. Navier-Stokes Equation does not have closed form solutions because of its chaotic nature. As for its numerical solutions, a general procedure is absent. 

Traditionally, the method of fluid dynamic optimization is for the experts to estimate based on empirical accounts and test potential possibilities with Wind Tunnel Testing (WTT). This approach is both costly and uncertain, as the prototyping involves the excessive use of building materials and the final result may not be proved the most optimized due to human incompleteness. 

New attempts to expedite the design include replacing WTT with computational simulations, using, for instance, MATLAB. But, still, the numerical solutions for Navier-Stokes are unstable and the drawbacks for empirical design remain. Automation attempts are represented by Bayesian optimization \citep{eismann2017optimization}. In this study, a Bayesian optimization model is applied to automate the design process. However, their model is sample-heavy, which leads to the undesirably exorbitant simulation cost.

To remedy these drawbacks, the presented study explores design automatic algorithms that design shapes with desired aerodynamic performance with reduced human input and computation consumption. Incorporating active learning ideas with DNN to map the shape and drag coefficient samples from input, in the form as latent dimensions, the model in this study is able to search for optimal shapes. A Frequentist active learning method facilitates the model fitting by focusing on only the potential output space regions, reducing the cost for producing excessive data points.

\textbf{This research differs from and builds on top of the previous studies in the following ways: }

\textbf{1. }MATLAB is utilized to form a new self-simulated dataset that includes various settings and object shapes, creating a template that can generate physical models that can reduce the costs of actually making desired objects and evaluating the attributes of them in reality. This approach is both less time-consuming and costly material-wise.

\textbf{2. }A new system of algorithm is raised to make the process of engineering optimization as automatic, and computational and sample economic as possible. DNN forms the core of the presented system with other self-determinant conditions to decide where the search should focus on. In such way, we make sure that the best fitting of the correlation between $\theta$’s and drags can be found, while other problems, such as local minimum, out-of-the-reasonable-range dilemma and so on, shall be avoided. 

\textbf{3. }The manner machine learning is applied in aerodynamic engineering optimization is new. The latest such work shows a different approach to the problem \citep{eismann2017optimization}. The simulation process is different in that a template to simulate the physical properties of certain fluid environments is targeted in thie research. We employ DNN that only gives moderate consideration to bias to find the correlation between $\theta$ and drag with a much smaller dataset, while still being able to predict on an comparably accurate basis. Other works are more engineering-based, not showing significant contributions to automate such industrial design.

\section{RELATED WORKS}

Fluid dynamics focuses on the study of moving air or fluid \citep{Dunbar2015B}. It investigates the interactions between fluid environment and solid object that moves through \citep{Timmermans2015M}. It is important because of its vital applications in fields such as aerospace engineering and vehicle production \citep{NASA2016Aero}. By studying the effects of fluid moving past a solid object, engineers can optimize their designs of aerodynamic machines \citep{Shieh2009Jyh-Cherng}. One core objective in this process is to minimize the drag of solid objects under a number of realistic restraints \citep{leeham2017drag}. However, the key equation, Navier-Stokes Equation itself that fluid mechanic works are based on has problems. Not only does it lack a set of closed-form solutions because of its chaotic nature, but the general rules of solving the equation numerically also does not exist \citep{FEFFERMAN}. Moreover, the equation remains one of the six mathematically unproven hypothesis listed by Clay Mathematics Institute \citep{Jaffe2000}. These controversies mean that experts have to design shapes based off of their experience and verify estimated shapes in a simulator or wind tunnel test (WTT) for multiple rounds. 

\begin{center}
\includegraphics[width = 0.45 \textwidth]{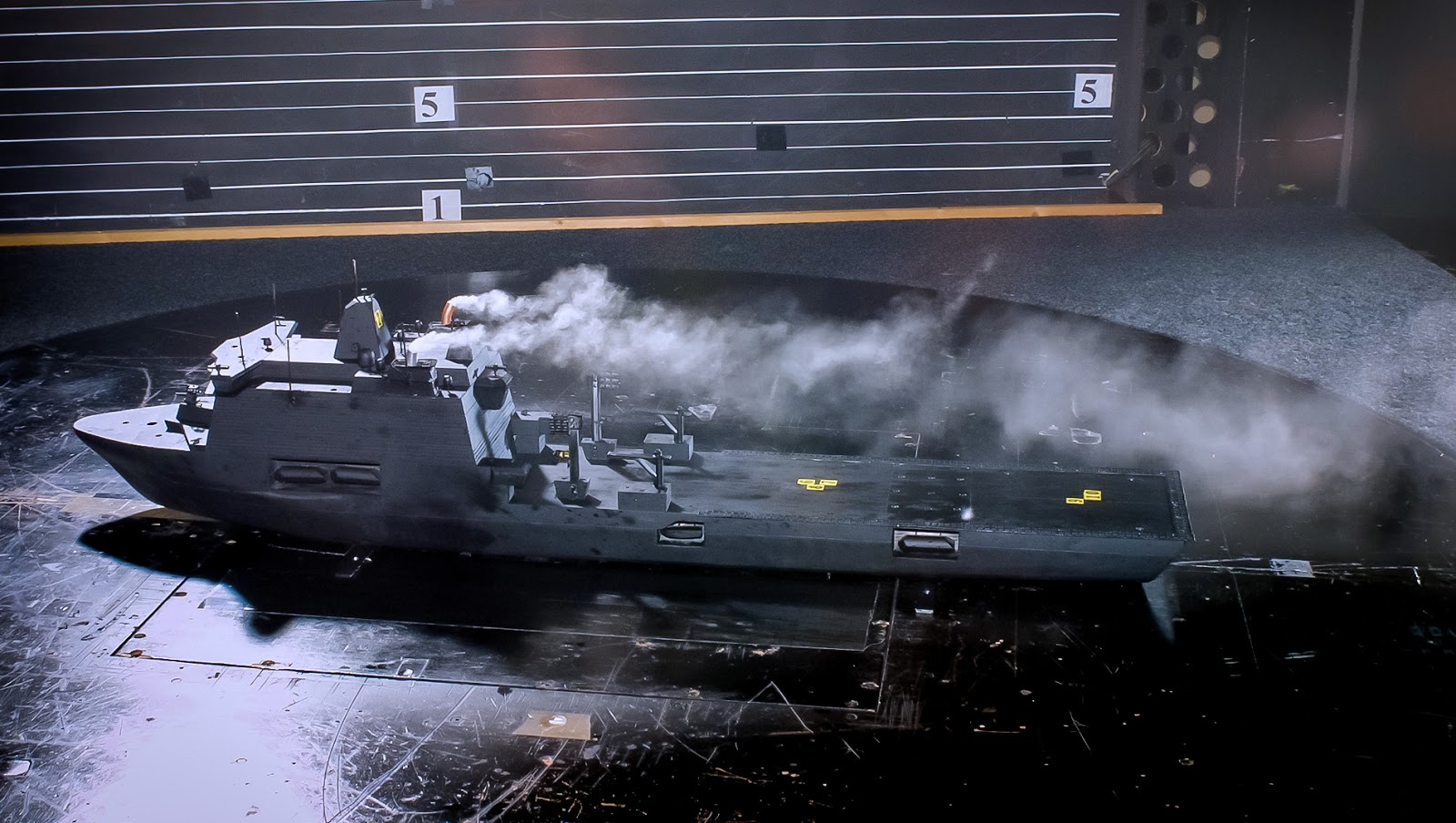}
Figure 1: Wind tunnel concept. A picture that illustrates a wind tunnel that tests the stream line design and many physical data of a ship \citep{WTC}
\end{center}

In a WTT, the subject of the test is placed in an open chamber. Examining how air propelled by a specialized fan would flow through the subject \citep{Richard2008Smith} pulls together a complete picture of the aerodynamic forces and other physical conditions on the model \citep{rossiter1964wtt}. Nevertheless, such a expensive method still requires expert knowledge. Results of such empirical estimations may not even be accurate after repeated prototyping. In general, the overall relationships between the shapes and drags are costly to search for under realistically given restrictions \citep{Bruce2016Jenkins}.

A group of researchers from Stanford University uses Bayesian optimization to accomplish this goal. Compared with the usual human search \citep{snoek2012bayesian}, the Bayesian model requires much fewer samples \citep{eismann2017optimization}. This late research in engineering optimization utilizes a statistical model to ease the complex enumeration process of samples while applying Bayesian optimization to find desirable aerodynamic design.

\begin{center}
\includegraphics[width = 0.45 \textwidth]{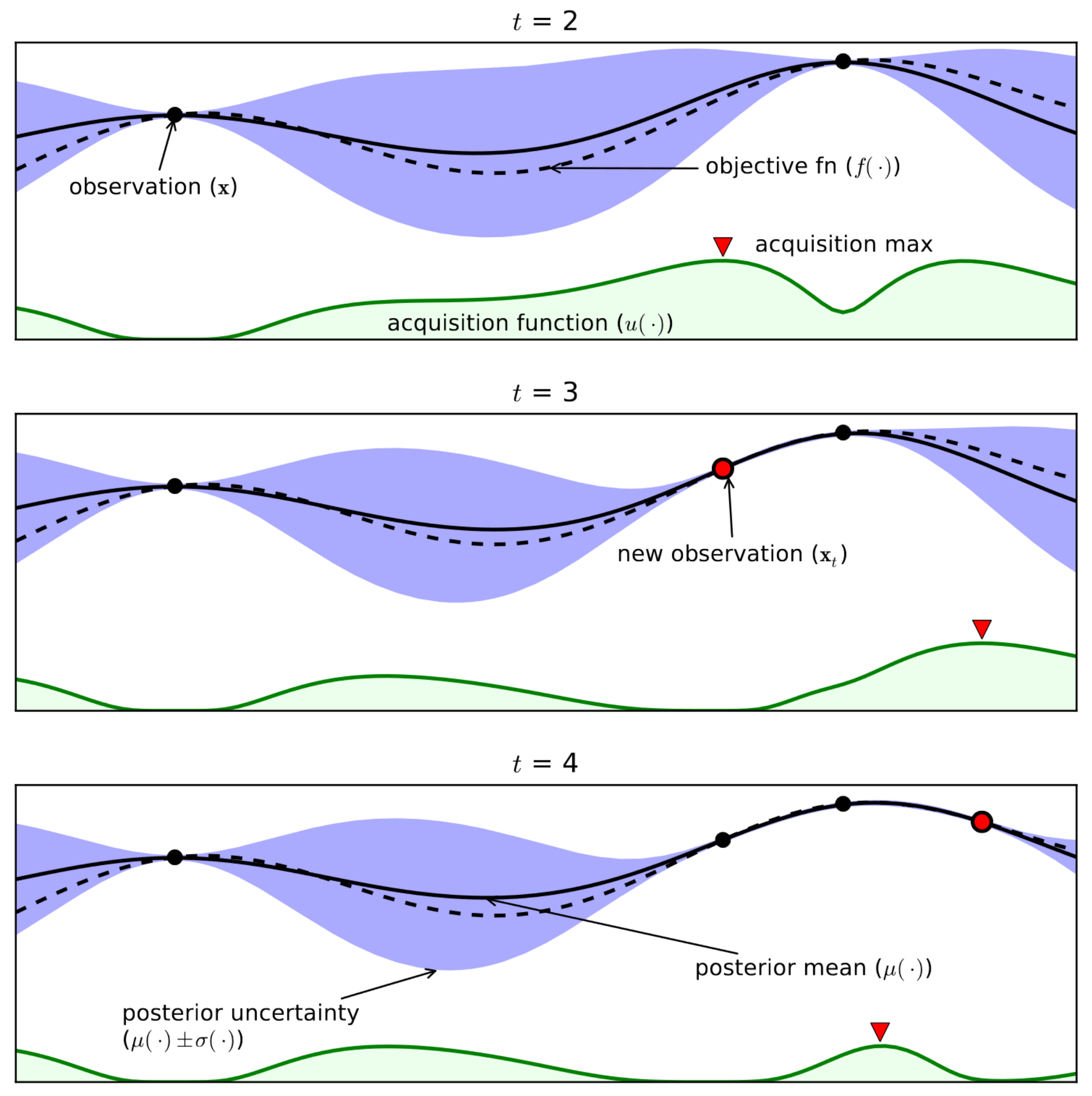}
Figure 2: Bayesian Optimization Fitting Process, Performed as a Form of Active Learning
\end{center}

Bayesian Optimization is also a symbolic representative of the application of active learning in Machine Learning \citep{martin1999pelikan}. It is by nature a Gausian method trained with parameter families based on $$ax+b, a \sim N(0, 1), b \sim N(0, 1)$$ As is seen in Figure 2 \citep{Puneith2017}. The regression is narrowed down sequentially with the addition of every new data point in a specific region that would provide the most advancement of regression fitting. This method significantly reduces the required number of data points by actively searching for desirable new input data for further training. However, Eismann's study is largely based on processing the simulated two dimensional images, which, because Bayesian model often shows an exponential increase of complexity with additional dimensions, requires a higher number of training data points to achieve complete training.

%% file: methodology.tex
\section{METHODOLOGY}



The core of our research is a DNN incorporated into a self-determinant loop that checks conditions and produces the optimized shape of the object under any given circumstance, which is set in MATLAB.  

\begin{center}

\includegraphics[width = 0.45 \textwidth]{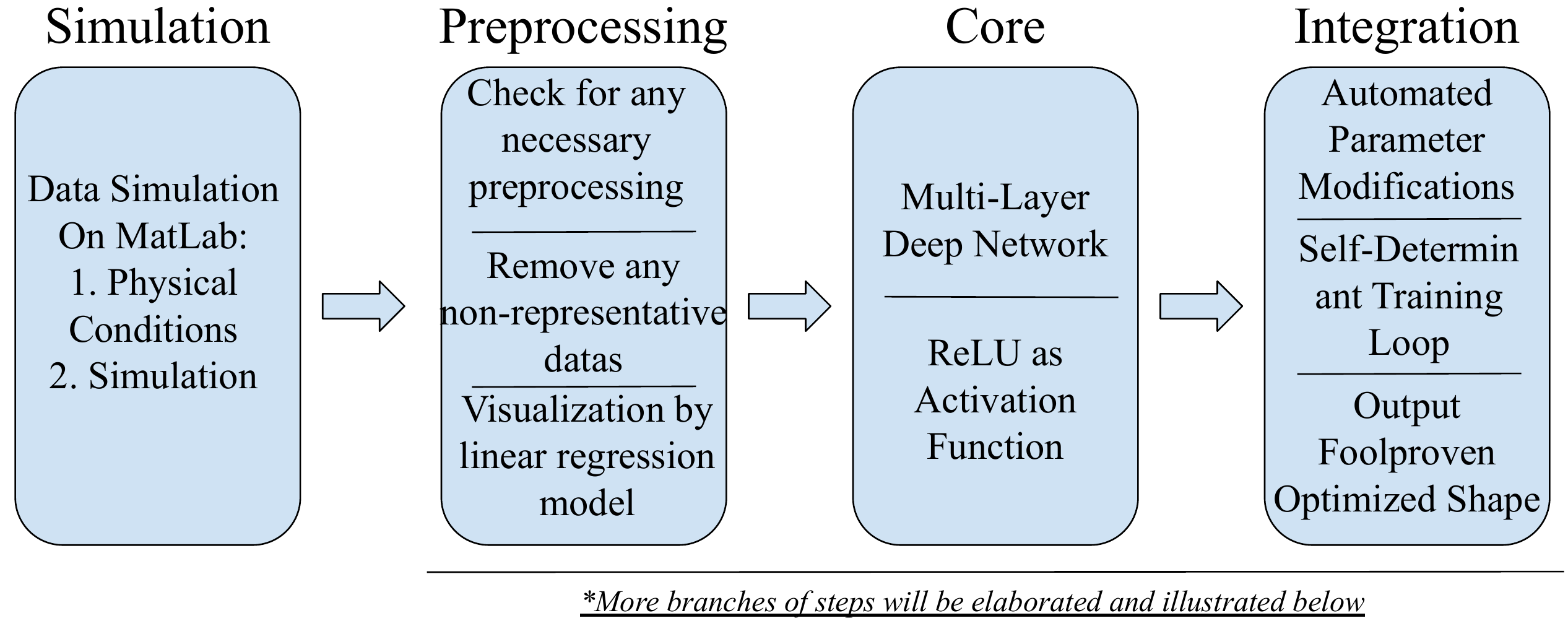}

Figure 3: Flow of the system, procedure outline. \textit{self-produced}.
\end{center}

Figure 3 shows the flow of the process. Our research focuses on making a comprehensive system that automatically finds an optimized shape under any given fluid environment. The process of aerodynamic optimization is simplified and made more digital based.

\subsection{Physical Simulation with MATLAB}

Resistances of objects of certain shapes in fluid, such as air and water, are studied in order for this research to be generalizable. The performance of individual design is defined by its drag coefficients ($C_D$) \citep{landau2013fluid}, which stand for the resistance an object encounters in fluid dynamic environment. $$C_D = \frac{F_D}{A \times \frac{\rho V^2}{2}}$$ For reference, $F_D$ stands for the drag force created by the fluid environment in the direction contrast the movement of the object. $C_D$ serves as the drag coefficient that depends on velocity, viscosity, and other parameters of the reference area. $A$, $\rho$ and $V$ are all environmental factors that reflect the reference area, density of the fluid and flow velocity relative to the object, respectively.

Incompressible flow in a volume satisfies the Navier-Stokes equation \citep{constantin1988navier}. $$\varrho((\frac{\partial u}{\partial t})+(u \cdot \nabla) = - \nabla p + v \nabla^2 u)$$

\textbf{In this equation:} 

\textbf{1. }$u$ represents the velocity of the flow, and is a vector field $V \mapsto R^3$ for 3-dimensional flow problems, or $V \mapsto R^2$ for 2-dimensional flow problems. 

\textbf{2. }$p$ represents the pressure in the volume, and is a function $V \mapsto R$. 

\textbf{3. }$v$ is the viscosity of the fluid (in this study, we set the viscosity to be \textbf{v = 0.2} as fluid water under room temperature to generalize our study).

The Navier-Stokes equation is the equivalent of Newton’s second law for fluids. To interpret the equation, we remark that $\frac{\partial u}{\partial t}$ the change in the flow with respect to time, and $(u \cdot \nabla)u$ represents the convective acceleration of the fluid. The right hand side represents the forces acting on the fluid: $\nabla^2 u$ is the difference between velocity of a point and the mean velocity of its neighborhood. This term encourages the vector field to become uniform in the absence of other influence factors. $\nabla p$ is the gradient of the pressure, and drives fluid motion. 

MATLAB is sued to simulate the dynamics of an object traveling through a fluid environment. To simulate the object, a geometry that represents the object and relevant boundary conditions is created. Then we convert the geometry into a mesh that represents the state of each point in the system. QuickerSim simulates fluid dynamics according to Navier-Stokes equation, and read off end results including drag through the provided API. Based on this setup, MATLAB generates a dataset of shapes and their corresponding drags.

\subsection{Fitting the Model}

\textbf{3.2.1 Linear Regression and Cross Validation} 

Linear regression is a linear combination of the features \citep{david2009statistical}. Linear regression expresses itself in the form below:
$$f(x)=\beta_{0} + \sum_{j=1}^{N} x_{j} \beta_{j}$$
It is also seen in the form as a matrix:
$$f(x) = a^{t}x$$
The loss function of such regression method is calculated in terms of the residual sum of squares, which is written in the expression as following:
$$RSS(\beta)=\sum_{i=1}^{n} (y_{i} - \beta_{0} - \sum_{j=1}^p x_{ij}\beta_{j})$$
This model is used first to approximate the correlation between the actual drag values. We choose to use such model because of its relatively low computational complexity as well as its ability to straightforwardly graph the relationship between actual and predicted drag values, if there is any. Later, by fitting our four-dimensional $\theta$ data again with DNN, we can directly visualize if there is any veritable improvement in our new model compared to the traditional ones.

\begin{center}

\includegraphics[width = 0.45 \textwidth]{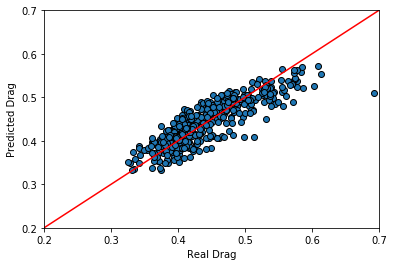}

Figure 4: Linear regression result compared to real drag values, while object width is set to 0.18. \textit{self-produced}.
\end{center}

Figure 4 represents how the linear regression result fits the real drag values. The closer all the blue dots are to the red curve, $f(x)=x$, the more accurate the linear regression prediction is. Some pattern is reflected on a primitive level as showns in Figure 4

Cross-validation is a statistic model, also known as rotation estimation, or out of sample testing \citep{Seymour1993}. The method tests the generalizability of certain predictions, which serves our purpose of the research well. In our research, the entire dataset of 625 $\theta$-drag pairs all directly serve as training data. Every four-dimensional $\theta$ has a corresponding drag label.

\begin{center}

\includegraphics[width = 0.45 \textwidth]{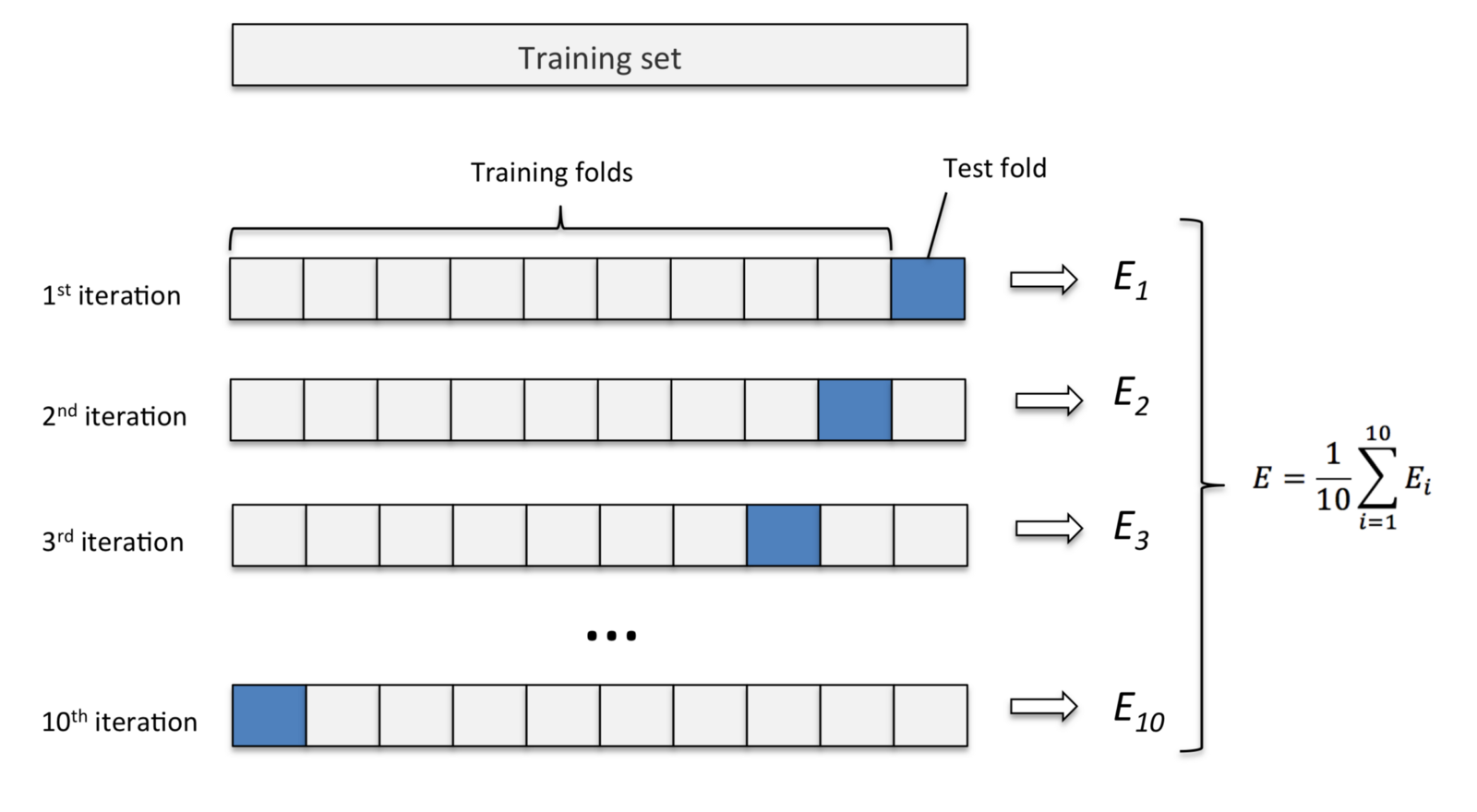}

Figure 5: cross-validation and resampling. \cite{karl2018kfold}
\end{center}

\textbf{3.2.2 Deep Neural Network (DNN)} 

Such a more expensive model as feed-forward DNN compared with other of the kind is suitable in our case, because our required computation is not as complicated as usual image processing tasks for two reasons. As is stated above in the MATLAB simulation explanation, we only require 625 samples to reach a fair prediction that is generalizable. This number is significantly smaller than that of digital image processing which usually desire around 10000 two-dimensional samples. Moreover, our input data consist of a list of arrays, which has a lower computational cost than two-dimensional images. Therefore, because of the remarkably lower computational cost as a nature of our dataset, DNN is feasible.

DNN is composed of Fully-Connected layer (FCL) which can be expressed in the following form: $$drag_{j}^{k} = f(\sum w_{ij}^{k}\cdot x_{j}^{k-1} + b_{j}^{k}) $$

$K$ represents a specific layer of the DNN, while $j$ represents the specific neuron the denoted variables are referring to. $W$ serves as a parameter that tells the connection between the $j^{th}$ neuron on $k^{th}$ layer and the $i^{th}$ neuron on $(k-1)^{th}$ layer, alongside $b$ as the bias adjuster. As the shape indicates, DNN can somewhat be regarded as a CNN in computation or thinking.

The formula can be extended to look something as the series of equations below:
$$Z^{(1)} =f^{(1)}(W^{(0)}X+b^{(0)}),$$
$$Z^{(2)} =f^{(2)}(W^{(1)}Z^{(1)}+b^{(1)}),$$
$$\cdots$$
$$Z^{(L)} =f^{(L)}(W^{(L-1)}Z^{(L-1)}+b^{(L-1)}),$$
$$Y(X) =W^{(L)}{Z^{(L)}+b^{(L)}},$$

While training, suitable activation functions is necessary. Rectified Linear Units (ReLU), represented by the expression below, is suitable in this model:
$$f(x)=max(0,x)$$
\citep{vinod2010relu}. ReLU is a fairly popular kind of activation function in neural network.

ReLU, a single-sided function, has a constant-valued slope when $x \geqslant 0$ so that it does not have sigmoids predicament of vanishing gradient. In the case of ReLU, only multiplications and comparisons are processed so that we may achieve a faster and more accountable convergence of results.

\subsection{Automated Search for Drag-Reduced Shapes}

The active learning (AL) used in this research is a Frequentist approach. It differs from Gaussian AL with an exponential increase of complexity in accordance with dimensions. In this study, the applied model maps the input space in a manner that only explores the specific regions indicated by DNN so that bias is traded in for a lower computation cost in the equations: $\mu = \dfrac1n \sum x_i$ (variance-significant); $\mu = 0$ (bias-significant), thus decreasing the required sample size to 625 data points.

\textbf{3.3.1 Restrictions for Self-Adjusting Parameters} 

Generally speaking, two parameters determine the way our model trains itself: train step and variable initializations. In the case of following conditions, our model would automatically determine to retrain itself: \textit{\textbf{1).} }when the taken derivative is found to be zero; or \textit{\textbf{2).} }the model is found to be stuck in a local minimum.

\textbf{3.3.2 Check for Decreasing Loss} 

After the first training round, there are two other possibilities, however, if not a single final optimized case is found, which are elaborated below:

\begin{algorithm}
\caption{Decreasing Loss Curve}\label{alg:euclid}
\begin{algorithmic}[1]
\State $\textbf{input}$ $dimension,$ \Comment{$m\times5$ $matrix$}
\State $drag$ \Comment{$m\times1$ $matrix$}
\State $\textbf{output}$ $optimized$ $step\_size$
\State $power \gets 1; STEP\_SIZE =\{\}; SCORE = \{\}$
\State $\textbf{while}$ $step\_size < 1$ $\textbf{do}$
    \State $\indent$ $step\_size \gets 10^{-6} \times 3^{power}$
    \State $\indent$ $STEP\_SIZE \gets STEP\_SIZE + \{step\_size\}$
    \State $\indent$ $power \gets power + 1$
    \State $\indent$ $L \gets $ $\{0, loss$ $per$ $1000$ $epochs\}$
    \State $\indent$ $score \gets 0$
    \State $\indent$ $\textbf{for}$ $i \gets 1$ $to$ $length.L$ $\textbf{do}$
        \State $\indent$ $\indent$ $\textbf{if}$ $L[i] > L\left[i-1\right]$ $\textbf{do}$
            \State $\indent$ $\indent$ $\indent$ $score \gets score + 1$
        \State $\indent$ $\indent$ $\textbf{end if}$
    \State $\indent$ $\textbf{end for}$
    \State $\indent$ $SCORE \gets SCORE + \{score\}$
    \State $\indent$ $\textbf{for}$ $element : SCORE$ $\textbf{do}$
        \State $\indent$ $\indent$ $temp \gets \infty$
        \State $\indent$ $\indent$ $\textbf{if}$ $element \leq temp$ $\textbf{do}$
            \State $\indent$ $\indent$ $\indent$ $temp \gets element$
            \State $\indent$ $\indent$ $\indent$ $holder \gets index$ $of$ $temp$
        \State $\indent$ $\indent$ $\textbf{end if}$
    \State $\indent$ $\textbf{end for}$
\State $\textbf{end while}$
\State $step\_size \gets STEP\_SIZE[holder]$
\State \textbf{return} $step\_size$
\end{algorithmic}
\end{algorithm}

Firstly, there may exist no reasonable loss convergence in our tested cases. If the loss does not decrease in a reasonable distribution or is not reduced at all, our system will start another round of training. In this round, the two train-step cases are le-2 and le-6, and so on until there is a reasonable convergence of loss. Figure 7 shows an example of how our model adjusts on its own in search of optimization:

However, in some cases, it is also possible that there is no liable convergence until it is out of the range for logical train step sizes. At such time, the system would automatically refresh its initializer so that there is a new initialization to avoid zero derivatives.

Secondly, too many reasonable convergences may be produced. On the one hand, to collate the performance accuracies, we juxtapose each of the test and train accuracies in each plausible training, and find out which convergence has the best general performance among all. On the other hand, to compare the stabilities, we take the difference between the level of test and train accuracies of each reasonable convergence respectively. The one that maintains relatively better train and test performances would be selected. Accuracies would be the priority factor of judgement.

\textbf{3.3.3 Check for Drag-Minimized Shape}

Undesirable initialization, incomplete fit of model and other factors unlisted may also lead to sub-optimal shapes. To prevent these situations, the mechanism in Figure 8 illustrates the test if the found optimized shape actually have the minimized drag in the given environment. Firstly, the found shape’s drag is the smaller than all existing shapes, it will be added to the original dataset and feedback to the training system for retraining. If the shape is actually also the global minimum in the new model, then it will be selected. Also, if the found optimized shape’s drag is not in fact the smallest compared to the previous drag values, then the found value will also be added to the training dataset. Our system would determine to get another round of retrain until a veritable drag-reduced shape is ensured.

\begin{algorithm}
\caption{Drag-Coefficient Minimization}\label{alg:euclid}
\begin{algorithmic}[1]
\State $\textbf{input}$ $min\_drag, min\_drag\;',$
\State $dimension$ \Comment{$m\times5$ $matrix$}
\State $drag$ \Comment{$m\times1$ $matrix$}
\State $\textbf{output}$ $min\_drag$
\While{$min\_drag\;' \geq min\_drag$}
\State $dimension = \{dimension + dimension \;'\}$
\State $drag = \{drag + drag\;'\}$
\State $\textbf{train}$ $f: dimension \to drag$
\State $\textbf{predict}$ $min\_drag\;'$
\EndWhile\label{euclidendwhile}\Comment{repeat one more time once false}
\State $min\_drag \gets min\_drag\;'$
\State \textbf{return} $min\_drag$
\end{algorithmic}
\end{algorithm}

\subsection{User Interface}

The final stage of the research centers around the implementation of a user interface that provides industrial designs based on boundary restriction given as input by the user and produce the optimal shape that fits.

\textbf{3.4.1 Boundary Function}

Industrial users may put in restrictions to produce an ideal shape for their specific requirements, including \textbf{viscosity} of fluid environment, and minimum body room \textbf{area}.

\textbf{Viscosity}
Viscosity is defined in MATLAB and the user input section is provided through Python interface. Without further ado, refer to 3.1 MATLAB simulation.

\textbf{Minimum Area}
To ensure that the result produced by the model offers enough enclose area to contain user-provided shapes, we apply \textbf{Flood Fill} algorithm as is visualized below in Figure 6 \citep{Torbert2016}.

With these judgement being made in the process, as the function, $$g: r \mapsto \{0, 1\}$$, that maps the vector direction of descent to either $0$, meaning the shape is in the boundary, or $1$, the shape is not in the boundary.

\begin{center}
\includegraphics[width = 0.45 \textwidth]{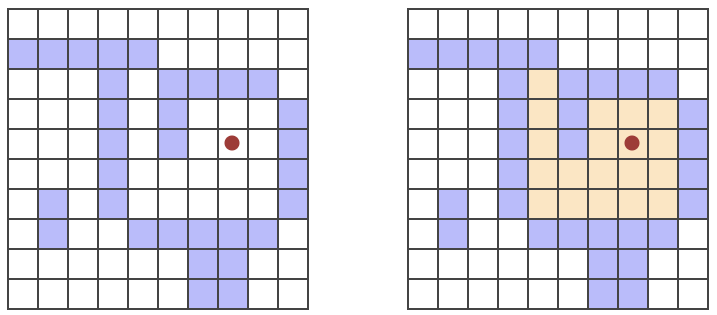}
Figure 6: Flood Fill Visualization \citep{flood2019}.
\end{center}

This algorithm assures the parameter surrounds a given inner shape. In reality, each square in the figure stands for a pixel: say, if the red dot is a user input, flood fill saturates the area around the dot so that there is no open section where the flood simply penetrates the boundary, which serves as a prerequisite for the minimum shape.

\textbf{3.4.2 Searching for the Optimal Shape}

\begin{center}
\includegraphics[width = 0.45 \textwidth]{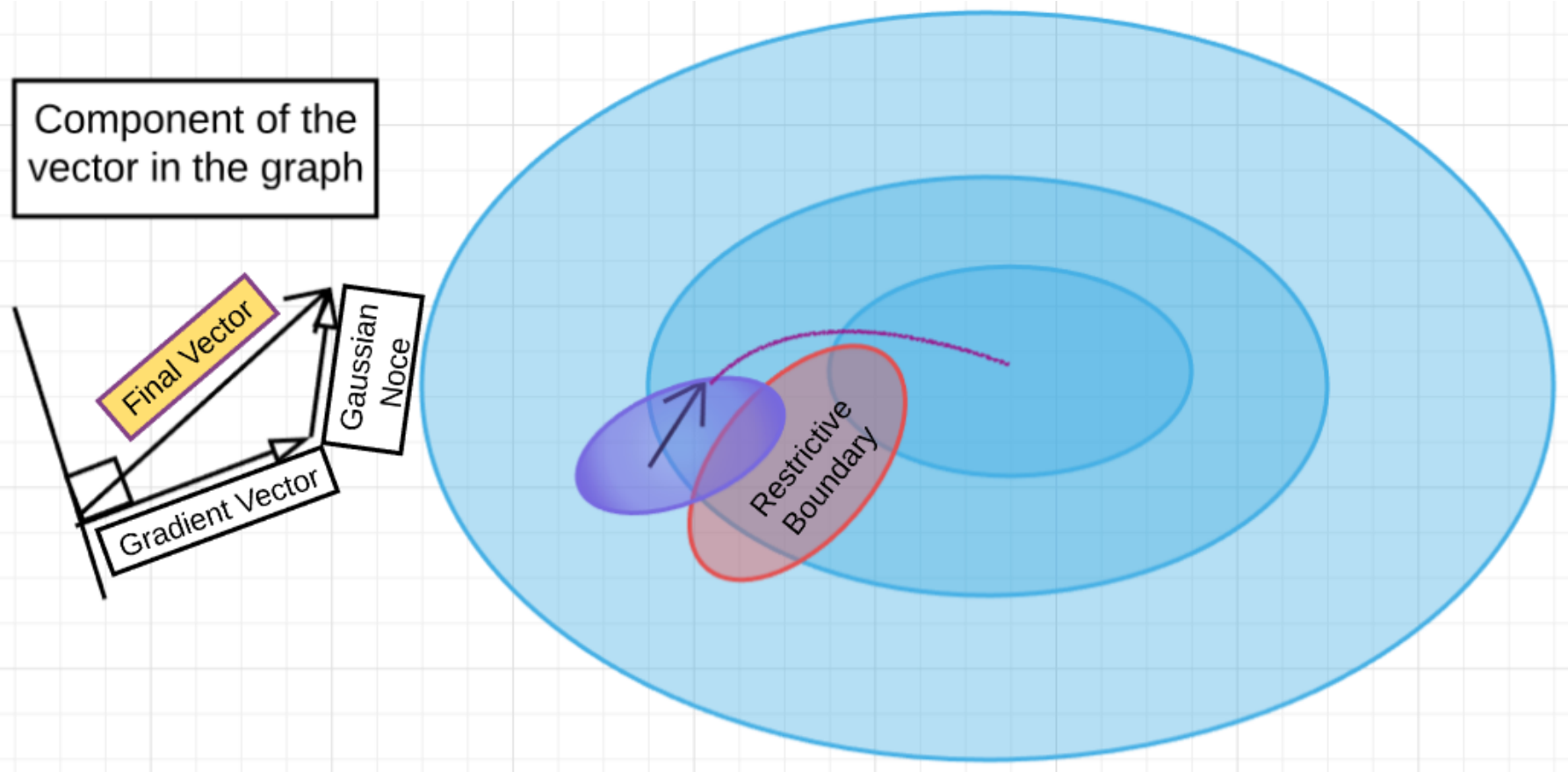}
Figure 7: Optimization with Restrictions, \textit{self-produced}
\end{center}

A method called \textbf{Stochastic Gradient Langevin Dynamics (SGLD)} is used to find optimization while successfully avoiding bypassing the restricted areas given by the user \citep{welling11}. The original \textbf{gradient descent} direction is given by $r' = r - \nabla _r L$. To avoid it to bypass the restricted area, we use a \textbf{half Gaussian noise} shown by $$r' + N(0, 1)$$ (it is based on a second order Gaussian/Normal Distribution: $n \sim N(0, \sigma^2)$) \citep{Rasmussen04}. It is half because it is limited only to form the vectors that point to directions $\leq 90^\circ$ to make sure the vector is descending, which is achieved by $$\arccos{\frac{(x,y)}{\parallel x \parallel \parallel y \parallel}} \leq 90^\circ$$.

The resulting path travelled by these vectors, will travel dynamically with a general path down the gradient direction.

%% file: results.tex
\section{EXPERIMENT}

\subsection{MATLAB Simulation}

We use MATLAB to create 3D environmental simulations and shadow the three-dimensional dynamic system into a two-dimensional graph and fit a surface.

\begin{center}
\includegraphics[width = 0.5 \textwidth]{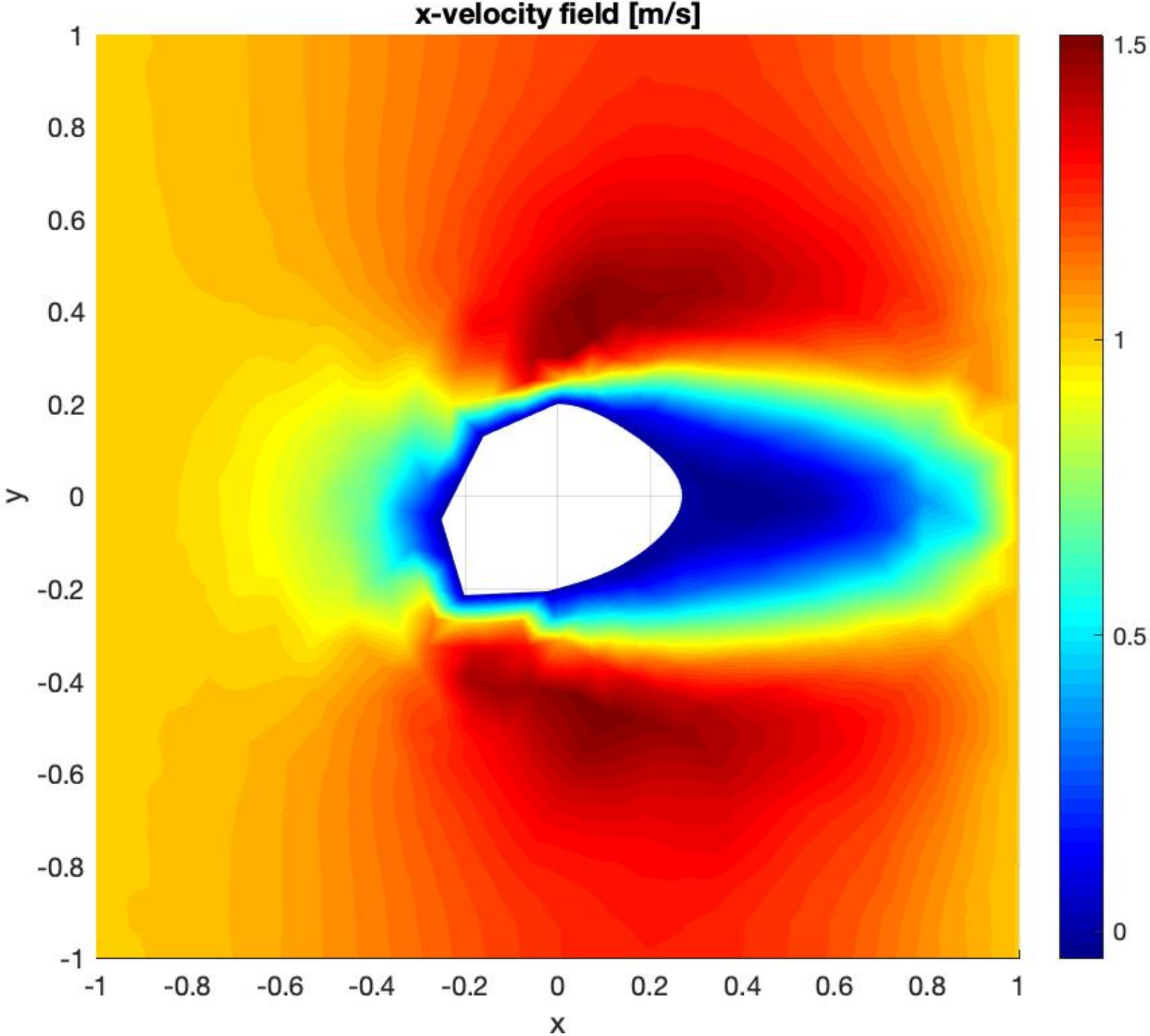}
\includegraphics[width = 0.4 \textwidth]{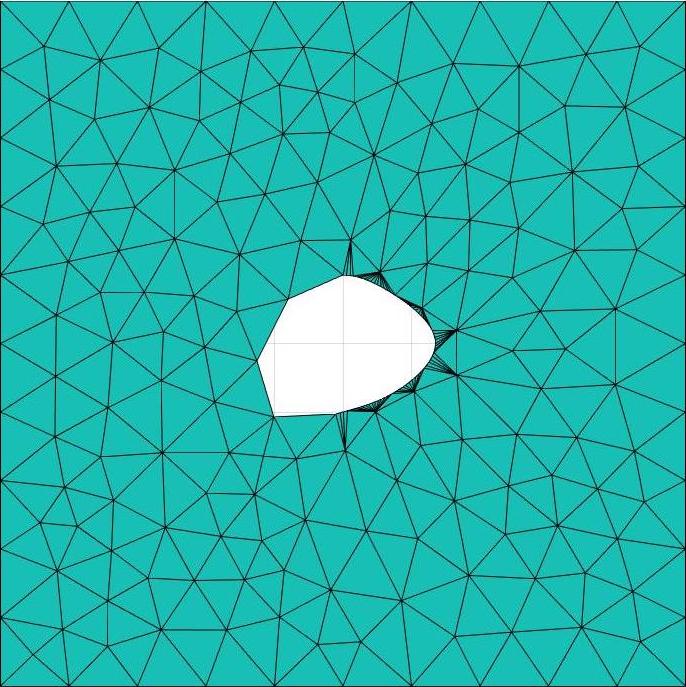}

Figure 8: MATLAB simulation examples \textbf{top:} Color-coded map represents relative velocity of environment against the object. \textbf{bottom:} Geometrical meshing of surrounding against moving objects. \textit{self-produced}.
\end{center}

In this case, only $625$ sample points is required for each case of our training. Dataset of such size can be simulated with MATLAB on a single local laptop in less than 3h (2.2 GHz Intel Core i7, 16 GB 2400 MHz DDR4).

As Figure 8 shows above, MATLAB template produces a simulation on a random, non-repetitive basis as such, along with their representing force if drag, for further training. B-SPLINE does well to visualize the boundary conditions of the 2D representation created.

\subsection{Deep Neural Network Regression}

Pre-training selection first takes place to eliminate simulations that is too close to singularity or off-scale to be for meaningful consideration. 

At this point, more accurate predictions of drag values from $\theta$’s are desirable. So, we develop a FCDNN built upon six FCL's which shows an optimized accuracy compared to the peers.

\begin{center}

\includegraphics[width = 0.45 \textwidth]{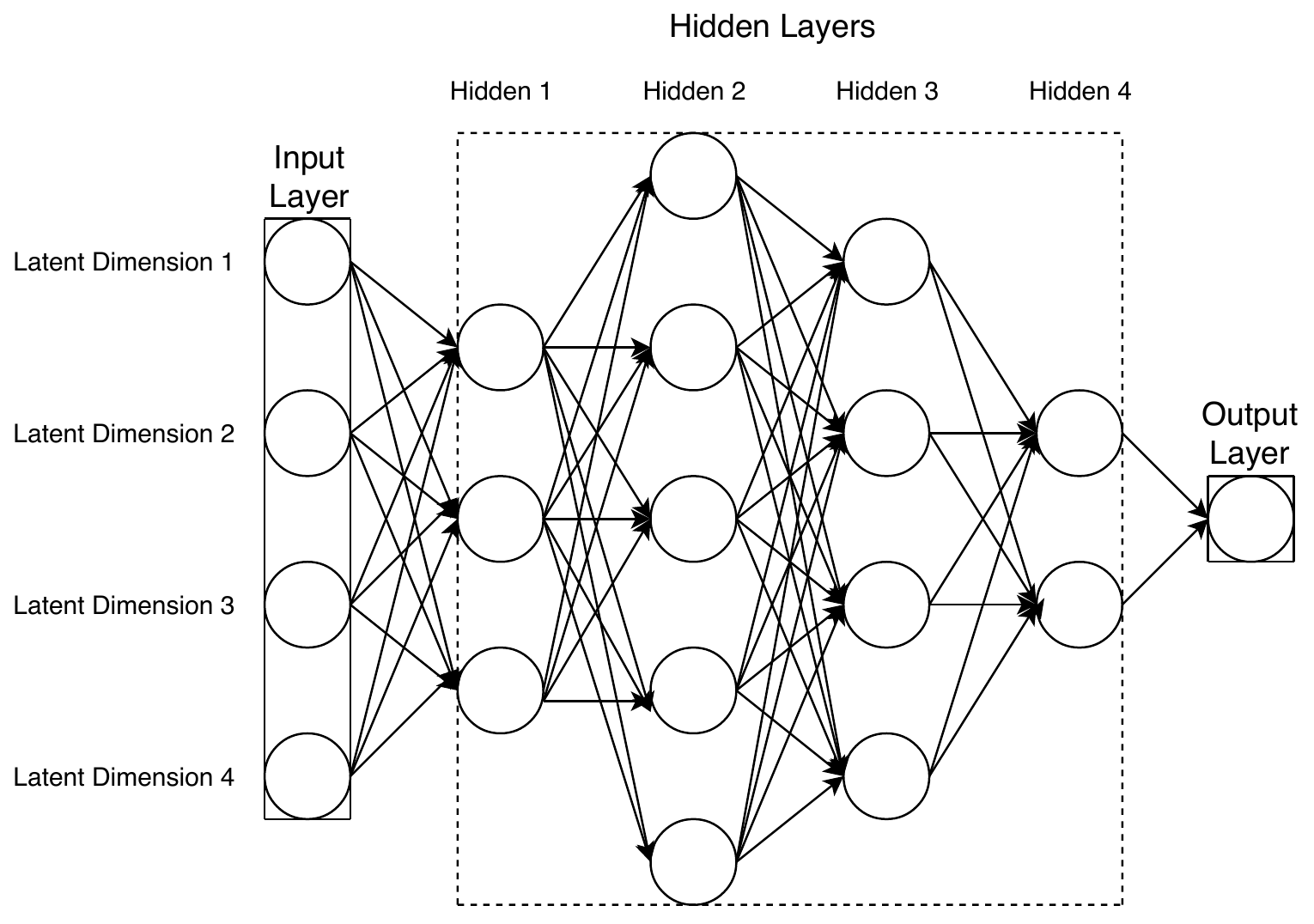}
Figure 9: Optimized DNN architecture in the case of 0.18 width. \textit{self-produced}.
\end{center}

Table 1 below shows examples of comparing performance between networks of different number of layers:
\begin{center}
\label{table:1} Table 1: Performance of DNN Architectures for Regression Loss

 \begin{tabular}{||m{1cm}||m{1.2cm}|m{1.2cm}|m{1.2cm}|m{1.2cm}||} 
 \hline
  Num. of Layers & Train Loss (0.15 Width) & Test Loss (0.15 Width) & Train Loss (0.18 Width) & Test Loss (0.18 Width) \\  
 \hline\hline
 Seven & $7.1E-4$ & $5.7E-4$ & $8.5E-4$ & $4.1E-4$\\ 
 \hline
 \textbf{\textcolor{orange}{Six}} & \textbf{\textcolor{orange}{$6.0E-4$}} & \textbf{\textcolor{orange}{$4.6E-4$}} & \textbf{\textcolor{orange}{$8.5E-4$}} & \textbf{\textcolor{orange}{$4.1E-4$}}\\
 \hline
 Five & $3.9E-3$ & $3.5E-3$ & $1.4E-3$ & $8.6E-4$\\
 \hline
 Four & $1.4E-3$ & $1.2E-3$ & $5.2E-2$ & $5.1E-2$\\
 \hline
\end{tabular}
\end{center}
With two width cases $\times$ four types of deep network architectures, eight different cases indicate where the optimized deep learning architecture can be achieved. As other different cases have been tested, there is a common trend shown by the two set of examples in Table 1. \textbf{Six-layer DNN} has significant improvement of performance in terms of MSE compared to other architectures with fewer layers, while another additional layer shows no perceptible reduction in MSE in the case of 0.18 width and actually increases MSE in the 0.15 width case. Therefore, six-layer DNN has superiority in its notably \textbf{low MSE} and \textbf{effectiveness}.

In both width cases and more, our AL over DNN model with its project-specific architecture is proven to make more accurate prediction in comparison with other traditional regression models, as is shown in Table 2:
\begin{center}
\label{table:2} Table 2: Performance of Different Models

 \begin{tabular}{||m{11em}||m{1.3cm}|m{1.3cm}||} 
 \hline
  & Linear Reg. & Our Model \\  
 \hline\hline
 0.15 Width Case Training & $8.1E-4$ & $2.9E-4$ \\ 
 \hline
 0.15 Width Case Testing & $7.6E-4$ & $1.9E-4$ \\
 \hline
 0.18 Width Case Training & $1.0E-3$ & $8.5E-4$ \\
 \hline
 0.18 Width Case Testing & $7.8E-4$ & $4.1E-4$ \\
 \hline
\end{tabular}
\end{center}
Table 3 visualizes the comparisons of computation costs of our study and a previous work based on Bayesian optimization. We reduce both simulation and training time by extracting dimensional information from 2D objects through the fitted spline and train with DNN model.
\begin{center}
\label{table:3} Table 3: Computation Cost Comparison with Related Study

 \begin{tabular}{||m{4.5em}||m{1.5cm}|m{1.1cm}|m{0.9cm}|m{0.9cm}||} 
 \hline
 Method & CPU & Memory & Simu- lation (hr) & Train- ing (hr) \\  
 \hline\hline
 Bayesian Optimization & Intel i7-3520M with 2.90 GHz & 16GB & 16 & 1 \\ 
 \hline
 \textbf{\textcolor{orange}{This Research}} & Intel i7-2400M with 2.20 GHz & 16GB & \textbf{\textcolor{orange}{4.5}} & \textbf{\textcolor{orange}{8E-2}} \\
 \hline
 Comparison &  &  & \textbf{\textcolor{red}{$72\%$ less}} & \textbf{\textcolor{red}{$92\%$ less}} \\
 \hline
\end{tabular}
\end{center}
Stacking of multiple different machine learning methods have insignificant boost to the accuracy. Stacking and ensemble function to remove unshared shortcomings of each method, which this study is not a suitable case. As for the better accuracy caused by the intrinsic advantage of other methods themselves, the method shows better performance in most cases through direct comparisons.

So with many factors taken into account, the six-layer FCDNN in this study is concluded an accountable model with \textbf{consistent performances} and \textbf{high level of accuracies} in both train and test. This architecture improves the loss of our prediction to be generally under 0.0005.

\subsection{Automated Engineering Optimization}

The Frequentist AL boosts the automation of our system. When the search of a drag-minimized shape falls into a local minimum or when the initialization of the training leads to a zero-derivative, the prediction result is not in its best case. Our system succeeds in avoiding such conditions.

\begin{center}
\includegraphics[width = 0.5\textwidth]{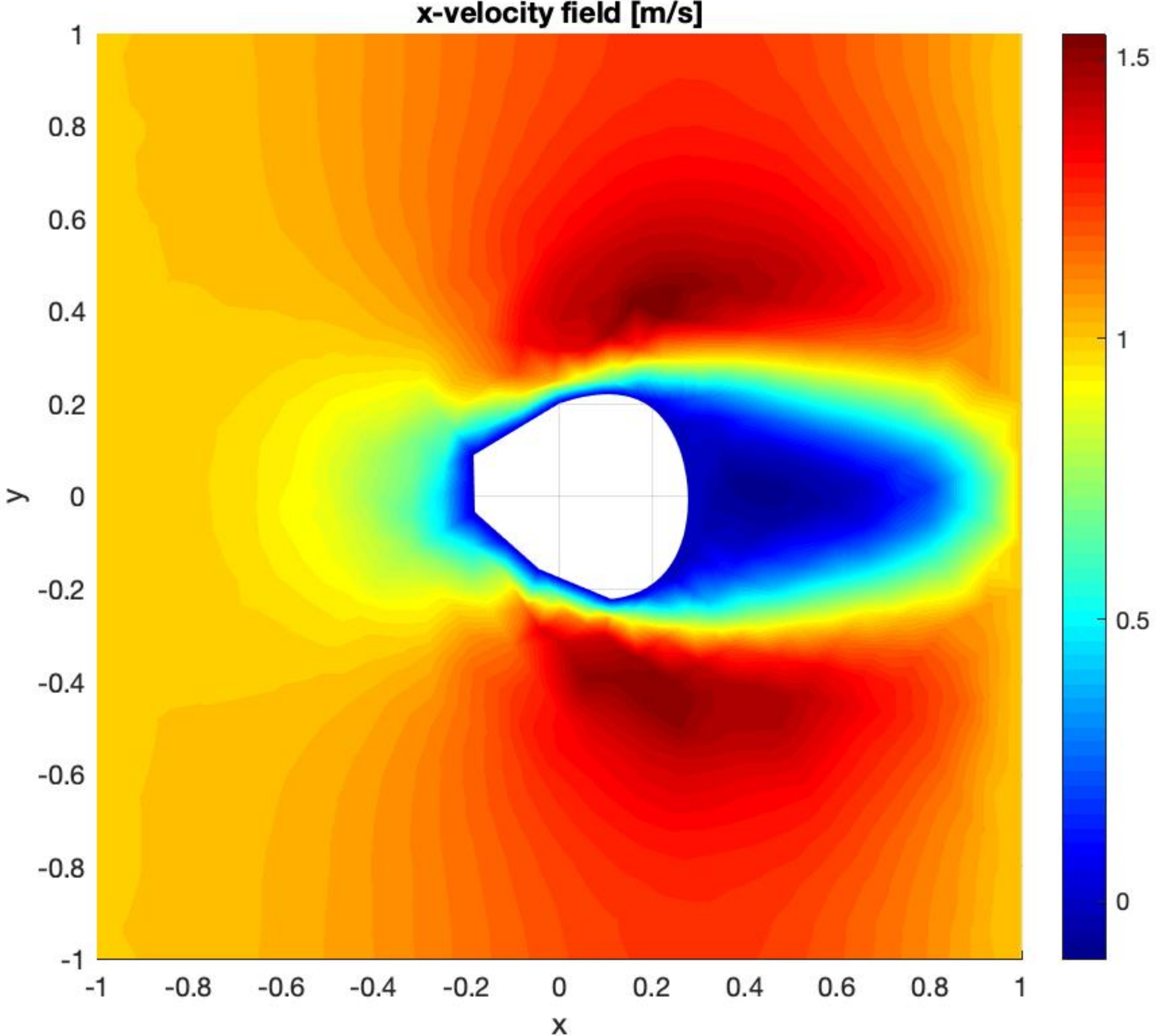}
\includegraphics[width = 0.4\textwidth]{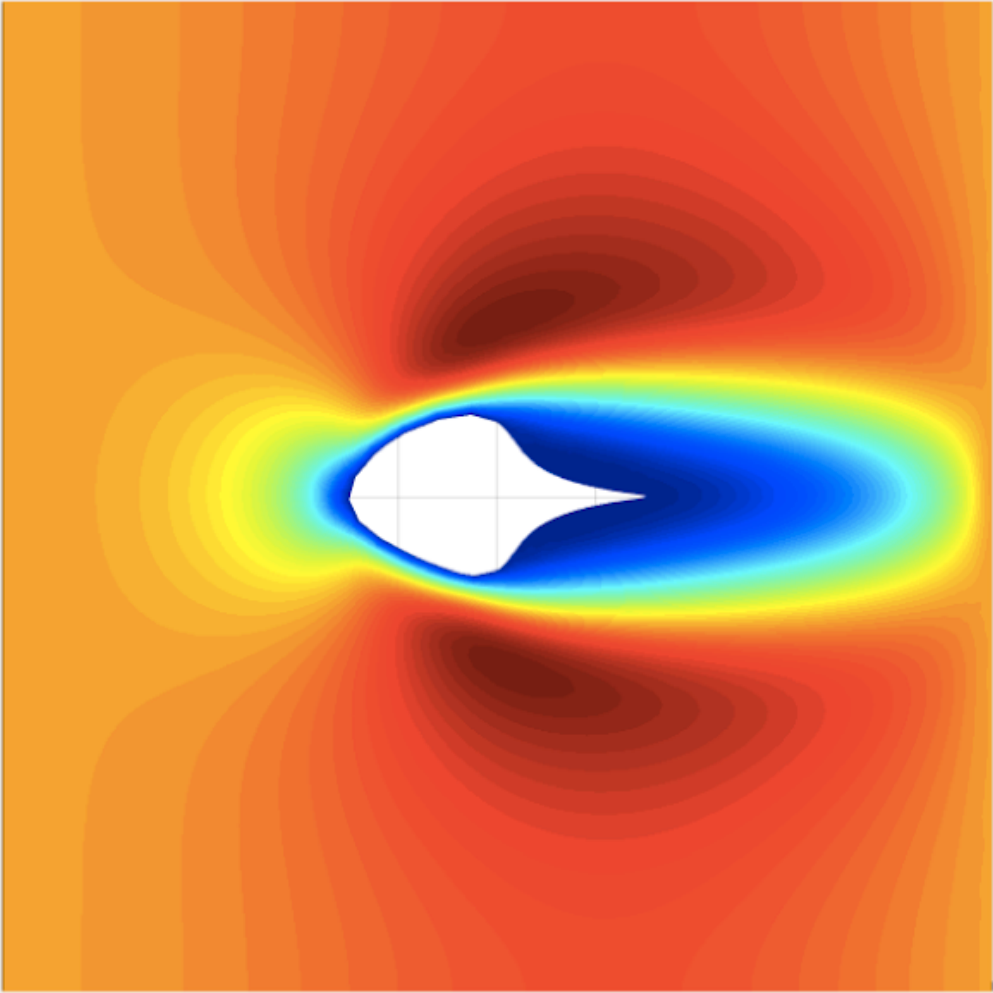}

Figure 10: Active learning advancement process \textbf{top:} Simulation results of the predicted optimized shape after first round training; \textbf{bottom:} The  predicted optimized result after the final round of training. \textit{self-produced}.
\end{center}

As shown in Figure 10, in this first effort to optimize shape, drag is shown to be \textbf{0.394} in Figure 10(a), which apparently is not indeed optimized and is captured by our mechanism. So it automatically goes into the second round of training, that reports the result below. After few rounds of active searches, our system automatically comes to the shape in Figure 10(b), which has a drag of only \textbf{0.281}, which is then proven to be the actual minimum. This reliability is similarly observed in other cases with widths of \textbf{0.18, 0.20, and 0.30}, which serves to support the efficacy of this method

%% file: discussion.tex
\section{DISCUSSION}

\subsection{Summary and Expectations}

The deep learning architecture in this research is new in its accuracy and stability performance. It perceptibly reduces the loss of prediction thus increase the efficiency. Visualized with the data above, this presented model is able to predict the relationship between $\theta$’s and drags accurately and precisely.

The application to use AI to find a fit for the correlation between $\theta$’s and drags is new to this research. We map a matrix of four-dimensional $\theta$ arrays to a column vector of one-dimensional drags, instead of finding the fit of the entire process of objects moving through given fluid environment. A relatively accurate result only requires a small data size.

This optimization system filters cases not yet fully trained and makes sure that the optimized shape is actually found with the simulated information. This also means that our system has the ability to avoid bugs in similar traditional algorithms.

\subsection{Proposed Future Extensions}

The following are potential aspects for our system to dig deeper into the field of work.

\textbf{5.2.1} So, supplementary parts may be used to boost the engineering performance of the shape. Consequently, a method of finding such appropriate add-on's may be studied to include into our system. Since we often need to set restrictions for shapes that we are optimizing to serve for specific tasks, the ultimate shape we get may not be close at all to the original global minimum of drag of our trained model.

\textbf{5.2.2} Other features may also be desired. For example, ability of an object maintaining its current height, stability or agility of an object. With varying purposes of the optimization, these conditions can be set into the MATLAB environment to be run through and tested the same way as we do on the drags.

%% file: conclusion.tex
\section{CONCLUSION}

In the end, our research successfully demonstrates that a more systematic and automatic aerodynamic engineering optimization is feasible by getting a regression mean error at below \textbf{0.0005} for most preset width cases and achieving actually drag-reduced shapes with the fit model through our systematic model. Compared to the previous researches, we have the advantages of requiring less training samples, less computation costs and time while improving the automation of engineering design and avoiding training bugs. On top of what we have already done, detailed additions as described in the discussion section is able to further increase the comprehensiveness of our system. These future bonus shall be achieved in a pretty similar way as we do here with the drags, except with different simulation features.

Our research successfully innovates on two things: finding a trend that relates to drag values, and searching for a drag-minimized shape. This provides a insight into how drag is influenced by objects’ shape with merely our machine learning prediction. This process can be more straightforward and practical than other methods that attempt to find such correlation.

%% file: Acknowledgments.tex
\begin{center}
\textbf{ACKNOWLEDGEMENTS}
\end{center}

The author would like to offer his cordial gratitudes for the assistance of the research mentor, Shengjia Zhao, on this research. He bears constructive advices to the author when challenges and confusions are to be dealt with.